\documentclass{elsart}

\usepackage{amssymb}

\newtheorem{remark}{Remark}

\newcommand{\tab}{\hspace*{2em}}
\newcommand{\algbox}[2][]{
  \vspace*{-0.5ex}
  \begin{center}
    \noindent
    \parbox{0.98\textwidth}{
      \noindent{#1}
      \framebox{\parbox{0.95\textwidth}{
          #2
        }}}
  \end{center}
  \vspace*{-0.5ex}
}

\renewcommand{\P}{{\mathcal P}}
\newcommand{\x}{\bar{x}}
\newcommand{\hx}{\hat{x}}

\newcommand{\opt}{\mbox{{\sc opt}}}
\newcommand{\optfrac}{\mbox{{\sc fopt}}}

\newcommand{\eps}{\varepsilon}

\begin{document}

\bibliographystyle{plain}
\runauthor{Kolliopoulos and Young}
\begin{frontmatter}

\title{Approximation Algorithms for Covering/Packing  Integer Programs}

\author[SGK]{Stavros G. Kolliopoulos}
\and 
\author[NEY]{Neal E. Young}

\address[SGK]{
Department of Informatics and Telecommunications, University of
Athens,  Panepistimiopolis Ilissia,  
 Athens 157 84, Greece (\texttt{www.di.uoa.gr/}\~{\tt sgk}).  
} 
\address[NEY]{
Department of Computer Science and Engineering, 
 University of California,
 Riverside, CA 92521, USA. 
 ({\tt neal@cs.ucr.edu}).}

\begin{abstract}
  Given matrices $A$ and $B$ and vectors $a$, $b$, $c$ and $d$, all
  with non-negative entries, we consider the problem of computing
  $\min \{c^{T} x : x\in \mathbb{Z}_+^n,\, Ax \ge a,\, Bx\le b,\, x \le d\}$.  %
  We give a bicriteria-approximation algorithm that, given
  $\eps\in(0,1]$, finds a solution of cost $O(\ln(m)/\eps^2)$ times
  optimal, meeting the covering constraints ($Ax\ge a$) and
  multiplicity constraints ($x\le d$), and satisfying
  $Bx\le(1+\eps)b+\beta$,  
where $\beta$ is the vector of row sums
  $\beta_i = \sum_j B_{ij}$. Here $m$ denotes the number of rows of
  $A.$ 
  
  This gives an $O(\ln m)$-approximation algorithm for CIP ---
  minimum-cost covering integer programs with multiplicity
  constraints, i.e., 
  the special case when there are no packing constraints $Bx\le b $.  The
  previous best approximation ratio has been 
  $O(\ln(\max_{j}\sum_{i}A_{ij}))$ since 1982.  CIP  contains the set
  cover problem as a special case, so $O(\ln m)$-approximation is the
  best possible unless P=NP.
\end{abstract}

\end{frontmatter}

\section{Introduction}
We consider integer covering/packing programs of the following form:

\medskip

{\em
Given $\P=(A,B,a,b,c,d)$ with
$A\in{}\mathbb{R}_{+}^{m\times n}$, 
$B\in{}\mathbb{R}_+^{r\times n}$,
$a\in{}\mathbb{R}_{+}^{m}$, 
$b\in{}\mathbb{R}_{+}^{r}$, 
and $c,d\in{}\mathbb{R}_{+}^{n}$,
compute
  \(\opt = \min\{c^{T} x : x\in \mathbb{Z}_+^n, Ax\ge a, Bx \le b, x\le d\}.\)
}

\bigskip

The constraints $Ax\ge a$, $Bx \le b$, and $x\le d$ are called,
respectively, {\em covering}, {\em packing}, and {\em multiplicity}
constraints.

The {\em width}, W, is $\min \{a_i / A_{ij} : A_{ij} > 0\}$.  Note
that it is easy to reduce any instance to an equivalent instance with
width $W$ at least 1 --- simply change each $A_{ij}$ to
$\min\{A_{ij},a_i\}$.  This does not change the set of integer
solutions.

The {\em dilation}, $\alpha$, is the maximum number of covering
constraints that any variable appears in.  

\medskip

A {\em $\rho$-approximate} solution is a solution meeting all constraints
and having cost at most $\rho$ times the optimum.  A {\em
  $\rho$-approximation algorithm} is a polynomial-time algorithm that
produces only $\rho$-approximate solutions. The quantity $\rho$ is
called the {\em approximation ratio} of the algorithm. 

\medskip

Perhaps the most well-known problem of the form above is set cover:
given a collection of sets with costs, choose a minimum-cost
collection of sets such that every element is in a chosen set.  
In the corresponding formulation 
$A_{ij} \in \{0,1\},$ and $a_i = 1,$ for $i=1,\ldots,m, j=1,\ldots,n.$  This
problem admits a simple $(1+\ln m)$-approximation algorithm
\cite{Johnson74,Lovasz75,Chvatal79}, and no $o(\ln m)$-approximation 
is possible in polynomial time, unless P=NP \cite{RazS97}.

Other special cases include natural generalizations of set cover,
including {\em set multicover} where  $a_i \in \mathbb{Z}_+$ 
and {\em multiset multicover} where in addition $A_{ij} \in \mathbb{Z}_+.$
\cite{Vaziranibook}.  In these problems, multiplicity constraints
limit the number of times a given set or multiset can be chosen.  In
facility-location problems (where $x_j$ represents the number of
facilities opened at a site $j$), multiplicity constraints are used to
limit the number of facilities opened at a site.  The motivation may
be capacity limits, security goals, or fault-tolerance (to ensure that
when a site is breached or damaged, only a limited number of opened
facilities should be affected) \cite{SrinivasanT97j,NaorR95}.

\medskip 

We give bicriteria approximation algorithms.
For any $\eps\in(0,1]$, our first algorithm finds a solution $\hx$ 
such that
\(A\hx\ge a, 
B\hx \le (1+\eps) b +\beta, 
\hx \le \lceil(1+\eps)d\rceil\), 
where $\beta$ is the vector of sums of
rows of $B$: $\beta_i = \sum_j B_{ij}$.  
The cost $c^T\hx$ is 
$O(1+\ln(1+\alpha)/(W\epsilon^2))$ times the optimum of the standard
linear programming (LP) relaxation.
Note that the standard
LP relaxation has an arbitrarily large integrality
gap if multiplicity constraints are to be respected.
Our second algorithm finds a solution $\hx$
of cost $O(1+\ln(1+\alpha)/\epsilon^2)$ times the optimum, 
satisfying 
\(A\hx\ge a, 
B\hx \le (1+\eps) b +\beta, 
\hx \le d\),
thus meeting the multiplicity constraints. 

These algorithms are appropriate for the case when $B$ has small row
sums (for example, a multiset multicover problem with restrictions
such as ``from the 5 sets $s_1,s_2,\ldots,s_5$, only 100 copies can be
chosen'') and for the  {\em CIP} (covering integer
programming) problem, formed by instances without packing constraints (no
``$Bx\le b$'').  CIP is  well-studied  in its own right.  For
this problem, our second algorithm is an
$O(\ln(1+\alpha))$-approximation algorithm.  This is the first
approximation algorithm for CIP whose approximation ratio is
logarithmic in the input size.  Fig.~\ref{fig:results} has a table of
known approximation
algorithms for CIP.\footnote{%
  In the table, $H(t)$ is the harmonic series with $t$ terms. It is
  well-known that $H(t) =
  \ln t + \Theta(1).$ To
  give some intuition for the Fisher-Wolsey bound consider for example
  the case where each $c_j = 1$ and the minimum non-zero entry of $A$
  is $1$.  In this case the bound is asymptotically equal to
  Dobson's.} 

\smallskip

\begin{figure}[h]
  \centering
\newcommand{\rr}[1]{{\raggedright{#1}}}
\begin{tabular}{|p{7.4em}||p{5.5em}|p{9.7em}|p{6em}|}
\hline
\rr{\bf who}&
\rr{\bf restriction\\on CIP}&
\rr{\bf cost approximation\\ ratio}&
\rr{\bf multiplicity\\guarantee}
\\ \hline\hline
\rr{Fisher\\ \& Wolsey \cite{FisherW82}}&
\rr{none}&
\rr{$1 +\ln(\beta_1/\beta_2)$
\\$\beta_1=\max_j {\sum_i A_{ij}}/{c_j}$ 
\\$\beta_2=\min  \{ \frac{A_{ij}}{c_j} | A_{ij} > 0 \}$}&
\rr{$x\le d$}
\\\hline
\rr{Dobson \cite{Dobson82}}&
\rr{$A_{ij}\in \mathbb{Z}_+$}&
\rr{$H(\max_{j=1}^n\sum_{i=1}^mA_{ij})$}&
\rr{$x\le d$}
\\\hline
\rr{Rajagopalan \\\& Vazirani \cite{RajagopalanV93}}&
\rr{$A_{ij}\in \{0,1\}$}&
\rr{$O(\ln (1+\alpha))$}&
\rr{$x\le d$}
\\\hline
\rr{Srinivasan\\ \& Teo \cite{SrinivasanT97j}}&
\rr{$c_j=1$}&
\rr{$O(1 + \ln(m) / (W\eps^2))$}&
\rr{$x\le \lceil (1+\eps) d\rceil$}
\\\hline
\rr{Kolliopoulos \cite{Kolliopoulos03}}&
\rr{$A_{ij}\in \{0,\phi_j\}$ \\for some $\phi_j$}&
\rr{$O(\ln (1+\alpha) )$}&
\rr{$x\le \lceil 12 d\rceil$}
\\\hline
\rr{Srinivasan \cite{Srinivasan95j,Srinivasan96}}&
\rr{none}&
\rr{$O(1 + \ln (1+\alpha) / W)$}&
\rr{\small$x\le O(1 + \ln (1+\alpha) / W) d$}
\\\hline\hline
\rr{this paper}&
\rr{none}&
\rr{$O(1+\ln(1+\alpha)/(W\eps^2))$}&
\rr{$x\le \lceil(1+\eps)d\rceil$}
\\\hline
\rr{this paper}&
\rr{none}&
\rr{$O(\ln(1+\alpha)/\eps^2)$}&
\rr{$x\le d$}
\\\hline
\end{tabular}
\caption{Approximation algorithms for the 
  CIP problem, \(\,\min\{c^{T} x : x\in \mathbb{Z}_+^n, Ax\ge a, x\le d\}\).
  The width $W$ is $\min\{ a_i/A_{ij} : A_{ij}>0\}$.  Without loss of
  generality, $W\ge 1$.  The dilation $\alpha$ is the maximum number
  of constraints any variable appears in.  The algorithms presented in
  this paper generalize to allow packing constraints ($Bx\le b$); for
  the general case the approximate solution $\hx$ satisfies $B\hx\le
  (1+\eps)b + \beta$ where $\beta_i = \sum_j B_{ij}$.  }
\label{fig:results}
\end{figure}

We use here results for another special case --- CIP without
multiplicity constraints.  This problem, which we denote CIP$_\infty$,
has a long line of research, but we use only the following results.
Randomized rounding easily yields an
$O(1+\ln(m)/W+\sqrt{\ln(m)/W})$-approximation algorithm, where $W$,
called the {\em width} of the problem instance, is
$\max\{a_i/A_{ij}:A_{ij}>0\}$.  
Srinivasan gives an
$O(1+\ln(1+\alpha)/W+\sqrt{\ln(1+\alpha)/W})$-approximation algorithm,
where $\alpha$, called the {\em dilation} of the instance, is the
maximum number of constraints that any variable occurs in
\cite{Srinivasan95j,Srinivasan96}.  Neither of these algorithms return
solutions that are suitable for CIP, as the solutions can violate the
multiplicity constraints by a large factor.

\medskip

A preliminary version of this paper appeared in
\cite{KolliopoulosY01}.  Other work on covering problems includes
\cite{Dobson82,FisherW82,RajagopalanV93,PlotkinST95,Young95,Srinivasan95j,Srinivasan96}.
See \cite{Hochbaum97chap} for a survey.

\medskip
The outline of this paper is as follows. In Section~\ref{sec::rho,l}
we present our first main algorithm that violates 
the multiplicity constraints by a $(1+\eps)$ factor. In
Section~\ref{sec::meet} we discuss 
the integrality gap of the standard LP formulation and present 
our second main algorithm which  meets the  multiplicity constraints.  
We conclude in Section~\ref{sec::discussion} with some open questions.

\section{Rounding LP relaxations of CIP$_\infty$ and CIP} 
\label{sec::rho,l} 

The approximation ratios in this paper are proven with respect to
various linear programming relaxations of the problems.  Our first
main result follows from careful consideration of the relation between
various forms of the problem and their standard relaxations.

\medskip

We begin by describing a standard approximation algorithm for
CIP$_\infty$.  Given an instance $\P=(A,a,c)$ of CIP$_\infty$, the
standard linear programming (LP) relaxation is
\(\optfrac_\infty = \min\{c^{T} x : x\in \mathbb{R}_+^n, Ax\ge a\}.\)
We call feasible solutions to this LP
{\em fractional} solutions to $\P$.
In contrast we call actual solutions to $\P$
{\em integer} solutions.

The value $\optfrac_\infty$ can be computed in polynomial time (using linear
programming) and is a lower bound on the optimum value $\opt$.  The
algorithm computes an optimal solution $\x$ (of cost $\optfrac_\infty$) to
the fractional relaxation, then rounds $\x$ to an integer solution
$\hx$ using the following randomized rounding scheme:
\begin{lem}[folklore]\label{lemma:folklore}
  Given a CIP$_\infty$ instance $\P=(A,a,c)$
  and fractional solution $\x$, 
  let $L=1+\max\{4\ln(2m)/W,\sqrt{4\ln(2m)/W}\}$.  With positive
  probability, the following rounding scheme produces an integer
  solution $\hx$ of cost at most $2L$ times the cost of $\x$:

  \algbox{\raggedright
      1.\ Let $x' = L \x$.
  
      2.\ Randomly round $x'$ to $\hx$:
      \\\tab let $\hx_j=\lceil x'_j\rceil$ 
      with probability $x'_j - \lfloor x'_j \rfloor$,  
      and $\hx_j = \lfloor x'_j \rfloor$ otherwise.
    }
\end{lem}
The proof is standard and we postpone it until the appendix. In what
follows the floor
(ceiling) of a vector $t$ denotes the vector where the $i$th
coordinate is the floor (ceiling) of $t_i.$ 

\begin{cor}
  Given a CIP$_\infty$ instance $\P=(A,a,c)$
  and fractional solution $\x$, 
  let $L=1+\max\{4\ln(2m)/W,\sqrt{4\ln(2m)/W}\}$.  One can compute in
  polynomial time an integer solution $\hx \le \lceil L\x\rceil$ of
  cost at most $2L$ times the cost of $\x$.
\end{cor}
The corollary follows because the rounding scheme can be derandomized
using the method of conditional probabilities 
\cite{ErdosS73,Raghavan88,Spencer87}.   
The rounding scheme
above has been improved by Srinivasan, who shows the following:
\begin{thm}[\cite{Srinivasan96}]  \label{theorem::sri} 
  Given a CIP$_\infty$ instance $\P=(A,a,c)$
  and fractional solution $\x$, 
  let $\alpha$ be the maximum number of constraints in which any
  variable appears.  
  For some $L=1+O(\ln(1+\alpha)/W+\sqrt{\ln(1+\alpha)/W})$,
  one can compute in polynomial time an integer solution
  $\hx\le \lceil L\x\rceil$ of cost $O(L)$ times the cost of $\x$.
\end{thm}
Since the optimal fractional solution $\x$ can be computed in
polynomial time, Srinivasan immediately obtains an $O(L)$-approximation
algorithm for CIP$_\infty$.

\subsection{Extending to CIP using $1/K$-granularity}

A natural idea would be to extend the rounding schemes above for
CIP$_\infty$ to handle CIP problems too.  Of course, to do this, we
need to figure out how to handle the multiplicity constraints.  The
natural LP relaxation of CIP is
\[\optfrac = \min\{c^{T} x : x\in \mathbb{R}_+^n, Ax\ge a, x\le d\}.\]
The first idea would be to compute the optimal fractional solution
$\x$, then use the rounding scheme from Lemma~\ref{lemma:folklore} or
Theorem~\ref{theorem::sri} to find an integer solution $\hx$
approximating $\x$.  But those rounding schemes return $\hx$ such that
$\hx\approx{}L\x$.  So, $\hx$ would violate the multiplicity constraints
by a factor of $L$.  But $L$ can be as large as $\Omega(\ln m)$, and
we would prefer to not violate the multiplicity constraints so much.

To work around this, given a CIP $\P=(A,a,c,d)$, we do compute an
optimal fractional solution $\x$, but then, instead of computing an
integer solution $\hx$ that approximates $\x$, we first compute a
fractional solution $\x'$ that is what we call {\em $(1/K)$-granular}
--- meaning that each coordinate of $\x'$ is an integer multiple of
$1/K$.  We do this for a sufficiently large integer $K$, so that the
$(1/K)$-granular solution $\x'$ has $\x' \approx (1+\epsilon)\x$ (and
satisfies all covering constraints).  To get the final integer
solution $\hx$, we round $\x'$ up deterministically by rounding each
coordinate up to its nearest integer.  Then $ \hx = \lceil \x'\rceil
\le \lceil(1+\epsilon)\x \rceil$.  A little thought shows that this
last rounding step increases the cost by at most a factor of $K$, so
that the cost of $\hx$ is $O(K)$ times the cost of $\x$.

The next lemma captures the exact tradeoff between granularity and
approximation of the cost (and, implicitly, multiplicity constraints).
The lemma is a straightforward consequence of the previous results.

\begin{lem}\label{lemma::granular} 
  Fix any integer $K>0$.  Given a CIP$_\infty$ instance 
  $(A,a,c)$ and
  fractional solution $\x$, let $\alpha$ be the maximum number of
  constraints in which any variable appears.  For some
  $L=1+O(\ln(1+\alpha)/KW+\sqrt{\ln(1+\alpha)/KW})$, one can compute in
  polynomial time a $(1/K)$-granular solution $x''\le \lceil L\x\rceil$ of
  cost $O(L)$ times the cost of $\x$.
\end{lem}
\begin{pf}
  Here is the algorithm.
  The input is  $\P=(A,a,c)$, $\x$, and $K$.

  \algbox{
      1.\ Construct CIP$_\infty$ instance $\P' = (A,Ka,c)$.
      Let $\x' = K\x$.
    
      2.\ Let $\hx'$ be the integer solution obtained by applying
      Theorem~\ref{theorem::sri} to $\P'$ and $\x'$.

      3.\ Return $x'' = \hx'/K$.  
    }
  Step 2 is well defined as $\x'$ is a fractional solution to
  $\P'$.  
  
  By Theorem~\ref{theorem::sri}, $\hx'\le \lceil
  LK\x\rceil$ is an integer solution to $\P'$ of cost $O(KL)$ times
  the cost of $\x$,
  with
  $L=1+O(\ln(1+\alpha)/KW+\sqrt{\ln(1+\alpha)/KW})$.
  
  Thus (using $x''=\hx'/K$), $x''\le \lceil L\x\rceil$ is a
  $(1/K)$-granular solution to $\P$ of cost $O(L)$ times the cost of
  $\x$.  (We also use here $\lceil LK\x\rceil/K\le \lceil L\x\rceil$
  for integer $K$.)
\qed \end{pf}
Note: In Step 2 of the algorithm in the proof,
Lemma~\ref{lemma:folklore} can be used instead of
Theorem~\ref{theorem::sri}, in which case the $1+\alpha$'s in the
definition of $L$ (in the lemma) are replaced by $m$'s.

\medskip
In the remainder of the section, by {\em a $(\rho,\ell)$-bicriteria
  approximate solution} for a CIP, we mean an integer solution $\hx$
that satisfies
$Ax \ge a$ and $x\le \lceil \ell d\rceil$,
with cost at most $\rho$ times the optimum $\optfrac$.  
By a {\em $(\rho,\ell)$-bicriteria approximation algorithm}, we mean a
polynomial-time algorithm that returns $(\rho,\ell)$-approximate
solutions.

Our first algorithm works as follows.  It first computes a
$(1/K)$-granular solution $\x'$ (where $K\approx\ln(1+\alpha)/(W\eps^2)$)
approximating the optimal fractional solution $\x$.  Then it gets an
integer solution $\hx$ by deterministically rounding each coordinate
of $\x'$ up to the nearest integer.  It returns $\hx$.

Here is a sketch of the analysis.  For this choice of $K$,
$\x'=(1+O(\eps))\x$, so that $\hx$ nearly satisfies the multiplicity
constraints: $\hx\le\lceil(1+O(\eps))\x\rceil$.  Since $\x'$ meets the
covering constraints, so does $\hx$.  Finally, $\x'$ has cost
$1+O(\eps)$ times the cost of $\x$, and, crucially, since $\x'$ is
$(1/K)$-granular, {\em deterministically rounding $\x'$ up increases
  the cost by at most a factor of $K$}.  So the final integer solution
$\hx$ has cost at most $K$ times the cost of $\x'$, i.e., $O(K)$ times
the cost of the original fractional solution $\x$.

The next lemma gives a detailed statement of the result and its proof.
\begin{lem}\label{lemma::bic}
  Fix any $\eps\in(0,1]$.  Given a CIP instance $(A,a,c,d)$ and fractional
  solution $\x$, one can compute in polynomial time an
  $(O(1+\ln(1+\alpha)/(W\eps^2)), 1+\eps)$-bicriteria approximate
  solution $\hx\le \lceil(1+\eps) \x\rceil$.
\end{lem}
\begin{pf}
  Here is the algorithm.  
  The input is  $\P=(A,a,c,d)$, $\x$, and $\eps$.

  \algbox{
    1.\ Take $K=\lceil \ln(1+\alpha)/ W \eps^2 \rceil$.
      
    2.\ Obtain a $(1/K)$-granular solution $\x'$ 
    by applying
    Lemma~\ref{lemma::granular} to the CIP$_\infty$ instance
    $\P'=(A,a,c)$ with fractional solution $\x$.
    
    3.\ Return $\hx = \lceil \x'\rceil$.  
  }
  By Lemma~\ref{lemma::granular}, for some
  $L=1+O(\ln(1+\alpha)/KW+\sqrt{\ln(1+\alpha)/KW})$, we have that
  $\x'\le\lceil L\x\rceil$ and that $\x'$ has cost $O(L)$ times the
  cost of $\x$.
  
  It follows that $\hx \le \lceil L\x\rceil$ and that $\hx'$ has cost
  $O(KL)$ times the cost of $\x$.  (The latter because $\x'$ is
  $(1/K)$-granular, which implies that the cost of $\hx$ is at most
  $K$ times the cost of $\x'$.)  Since (by the choice of $K$)
  $L=1+O(\eps)$, this implies the result.
\qed \end{pf}
\begin{remark}
The result of the lemma is best possible in the following sense.
For any finite $\rho,$ a $(\rho,1)$-approximate solution 
w.r.t.\ \optfrac  \ is impossible because of the arbitrarily large 
integrality gap (see Section~\ref{sec::meet} for an example).  
It is also well-known that the integrality gap for \ $\optfrac_\infty$  \   
is  $\Omega (\ln m)$  for the special case of set cover 
where 
arbitrarily large values for the variables are  allowed. Hence for any
$l,$  a $(\rho,
l)$-approximate solution for a CIP with $\rho = o (\ln m)$ is also impossible. 
\end{remark}

Now we can state our first main result --- an approximation algorithm
for any general integer covering/packing problem with multiplicity
constraints:
\[\opt = \min\{c^{T} x : x\in \mathbb{Z}_+^n, Ax\ge a, Bx \le b, x\le d\}.\]
The algorithm returns a solution that meets the covering constraints,
approximately meets the multiplicity constraints
(and hence approximately meets the packing constraints),
and has cost $O(K)$ times the cost \ \optfrac \ of the fractional solution.

\begin{thm}[first main result]\label{theorem::bic}
  Let $\eps\in(0,1]$,
  and an integer covering/packing program
  \(\opt = \min\{c^{T} x : x\in \mathbb{Z}_+^n, Ax\ge a, Bx \le b, x\le d\}\),
  with fractional solution $\x$,
  be given.
  Let $\beta_i = \sum_j B_{ij}$.
  Then one can compute in polynomial time an $\hx\in \mathbb{Z}_+^n$ such that

  \(
  \begin{array}{lrclcl}
    1. & c^{T} \hx & \le & O(1+\ln(1+\alpha)/(W\eps^2)) \,c^{T} \x
    \\2. & A\hx 
    & \ge & A\x 
    & \ge & a,
    \\3. & \hx  
    & \le & \lceil(1+\eps)\x\rceil
    & \le  &  \lceil(1+\eps)d\rceil, \mbox{ and }
    \\4. & B\hx
    & \le &(1+\eps)\x+\beta
    & \le &(1+\eps)b+\beta.
  \end{array}
  \)
\end{thm}
\begin{pf}
  Here is the algorithm.  
  The input is  $\P=(A,B,a,b,c,d)$, $\x$, and $\eps$.

  \algbox{
      1.\ Let $\hx$ be the approximate
      solution obtained by applying Lemma~\ref{lemma::bic}
      to the CIP instance $\P' = (A,a,c,d)$,
      and fractional solution $\x$.
    
      2.\ Return $\hx$.
    }

  Properties 1-3 of $\hx$ follow immediately from Lemma~\ref{lemma::bic}.
  To see that property 4 holds, note that,
  from $\hx\le \lceil (1+\eps)\x\rceil$ it follows
  that $\hx_j < (1+\eps)\x_j + 1$, which implies
  \((B\hx)_i \le (B(1+\eps)\x)_i + \beta_i\).
\qed \end{pf}

The optimal fractional solution $\x$ to the LP relaxation can be
computed in polynomial time, so Theorem~\ref{theorem::bic} immediately
implies that the desired approximate solution $\hx$ (having properties
1-4 from the theorem and cost $O(1+\ln(1+\alpha))\optfrac$) can be
computed in polynomial time.

\begin{remark} Note that for a CIP problem
with $\max_j d_j = O(1)$, by taking $\eps=1/(2\max_j d_j)$, the
above theorem implies that one can find in polynomial time an integer
solution having cost $O(1+\ln(1+\alpha)/W)\optfrac$ and $\hx_j\le
d_j+1$.  That is, the multiplicity constraints can be met within an
{\em additive} 1.
\end{remark}

\section{Meeting the multiplicity constraints}
\label{sec::meet} 

Given a fractional solution $\x$, it is not in general possible to
find an integer solution $\hx$ meeting the covering and multiplicity
constraints exactly and having cost $O(\ln (1+\alpha))$ times the cost of
$\x$.  To see this, fix $\delta>0$ arbitrarily small, and consider the
following CIP, which is a simple instance of Minimum Knapsack:
\[\min\{x_2 : x\in \mathbb{Z}_+^2, (1-\delta)x_1 + x_2 \ge 1, x_1\le 1\}.\]
The optimal fractional solution has cost $\delta$, whereas the optimal
integer solution has cost $1$.  This example demonstrates that the
integrality gap can be arbitrarily large if multiplicity constraints
are to be respected.\footnote{A 
similar example appears in 
\cite{CarrFLP00}.  In \cite{RajagopalanV93} the integrality
  gap was erroneously claimed to be
  $H(\max_{j=1}^n\sum_{i=1}^mA_{ij}).$  } 
However, notice
that the two constraints ($(1-\delta)x_1 + x_2 \ge 1$ and $x_1 \le 1$)
imply a third: $x_2 \ge \delta$.  This third constraint, and the
observation that $x_2\in \mathbb{Z}$, imply $\delta x_2 \ge \delta$.  

The constraint ``$\delta x_2 \ge \delta$'' above is a {\em valid
  inequality} for the CIP, meaning that it holds for all feasible
integer solutions.  Adding a valid inequality  to the integer program (IP)
does not change the space of solutions or the value of the optimal
solution.  But adding the constraint can strengthen the linear
programming relaxation by ruling out some fractional solutions, and
this can give a better bound on $\opt$.  For example, adding the
constraint to the example above, and then solving the LP relaxation
with the added constraint, gives a lower bound of 1 on $\opt$.

For the general problem, reasoning as above leads to a class of valid
inequalities called Knapsack Cover (KC) inequalities.  These
inequalities generalize valid inequalities used for CIP problems 
with $A_{ij} \in \{0,1\}$  in
\cite{Balas75,HammerJP75,Wolsey75}.  They were also used by Carr et al.
\cite{CarrFLP00}.

Our next algorithm begins by finding a fractional solution $\x$ to the
LP relaxation with a number of KC inequalities added.  It then rounds
$\x$ to an integer solution $\hx$ as follows: for $j$ such that $\x_j
\ge d_j/(1+\eps)$, it ``pins'' $\hx_j = d_j$.  (This increases the
cost by at most $1+\eps$.)  To set the remaining $\hx_j$'s, it rounds
the corresponding $\x_j$'s using the randomized rounding algorithm
from (Lemma~\ref{lemma:folklore}) or Srinivasan's algorithm
(Theorem~\ref{theorem::sri}).  Since each non-pinned $\x_j$ is at most
$d_j/(1+\epsilon)$, this rounding can be done so that $\hx_j$ is at
most $d_j$.

An astute reader may ask whether this process will work if started
with a fractional solution $\x$ to the LP relaxation {\em without} KC
inequalities.  If so, this would yield a faster algorithm.  After we
describe and analyze the algorithm sketched above, we discuss this
question.

\medskip

\subsection{The KC inequalities}
Fix a problem instance $\P=(A,B,a,b,c,d)$.  For each constraint
$(Ax)_i \geq a_i$ and any subset $F$ of the $j$'s (corresponding to $x_j$'s
that we imagine pinning), define
$a_i^F \doteq \max\{0, a_i - \sum_{j \in F} A_{ij}d_j \}$.
Define also $A^F_{ij} \doteq \min \{A_{ij},a_i^F \}$ for $j\in F$
and $A^F_{ij} \doteq 0$ for $j\not\in F$.
In words, $a_i^F$ is the residual covering
requirement of the $i$th constraint if all variables in $F$ were to be
set to their upper bounds, and $A^F_{ij}$ is $A_{ij}$, possibly lowered
to ensure the width is at least $1.$ 
(In the small example above, we knew that, for
$x_2\in \mathbb{Z}_+$, the inequality $x_2\ge\delta$ held if and only if the
inequality $\delta x_2 \ge \delta$ did, so we replaced the former
constraint with the latter.)
The KC inequalities for a set $F \subset N$ are 
\(A^F x \,\geq\, a^F\).
The {\em LP-KC relaxation} of $\P$ is to find $x\in \mathbb{R}_+^n$ minimizing
$c^{T} x$ subject to $Ax \geq a$, $Bx\le b$, $x\leq d$, and subject to
the KC inequalities for all $F \subset N$.

We are not aware of an algorithm that solves this relaxation exactly
in polynomial time. Carr et al.\ \cite{CarrFLP00} define the following type of
solutions, which are adequate for our purpose.  For
$\lambda >1$, call a vector $x$ a {\em $\lambda$-relaxed solution} to
LP-KC if it has cost at most the fractional  optimum of LP-KC and
satisfies (i) $Ax \geq a$, (ii) $Bx\le b$, (iii) $x \leq d$ and (iv)
the KC inequalities for the set $F_\lambda=\{j:x_j\geq d_j/\lambda\}$.
The following theorem follows from  the results in \cite{CarrFLP00} together 
with the properties of the ellipsoid method (see, e.g., \cite{Lovasz86}).
\begin{thm}[\cite{CarrFLP00,Lovasz86}]  \label{theorem::ellipsoid} 
  Suppose $\P=(A,B,a,b,c,d)$ has rational coefficients.  For any
  constant $\lambda >1$, a $\lambda$-relaxed solution to the LP-KC relaxation
  of $\P$ can be found in polynomial time.
\end{thm}
For the sake of completeness we sketch the idea behind the theorem. 
When the ellipsoid method queries the separation oracle with a point $x,$
the oracle  returns
a separating hyperplane corresponding either to a constraint of 
the  standard LP, or to one that is a 
valid KC inequality for the set of variables in $x$ that are high 
(in this particular $x$).  In the end, 
look at the set of hyperplanes the separation oracle has passed to the
ellipsoid method.
That set defines a polytope which  is a relaxation of the LP-KC polytope.

The input to our next algorithm is 
an instance $\P=(A,B,a,b,c,d)$ of the general
problem
and an
$\epsilon\in(0,1]$. The algorithm can also be viewed as a reduction of
  the problem of 
  finding a $\rho$-approximate solution to a CIP to finding a $(\rho,
  \ell)$-bicriteria approximate solution for appropriate $\ell.$
\algbox{
1.\ Set $d':=\lfloor d\rfloor$.

2.\ Let $\x$ be a $(1+\eps)$-relaxed solution to the LP-KC relaxation of
$\P = (A,B,a,b,c,d').$

3.\ Let $F=\{j : \x_j \ge d'_j/(1+\eps)\}$. 

4.\ Define CIP $\P'= (A',a',c,d'')$  by setting $A' := A^F,$ $a' :=
a^F,$  and defining fractional solution $\x'$ and $d''$ 
as follows:

5.\ For $j\in F$ let $\x'_j = d''_j = 0$.
For $j\not\in F$ let $\x'_j = d''_j = \x_j$.

6.\ Find integer solution $\hx'$ to $\P'$ by applying
Theorem~\ref{theorem::bic} with fractional solution $\x'$ and
the given $\eps$.

7.\ Let $\hx_j = d_j$ for $j \in F$ and $\hx_j = \hx'_j$ for $j\not\in
F$.  Return $\hx$.  }

\begin{thm}[second main result]   \label{theorem::reduction} 
  Given $\eps\in(0,1]$,
  and an integer covering/packing program
  \(\opt = \min\{c^{T} x : x\in \mathbb{Z}_+^n, Ax\ge a, Bx \le b, x\le d\}\),
  let $\beta_i = \sum_j B_{ij}$.
  The algorithm above computes in polynomial time
  an $\hx\in \mathbb{Z}_+^n$ such that

  \(
  \begin{array}{lrcl}
    1. & c^{T} \hx & \le & O(1+\ln(1+\alpha)/(W\eps^2)) \opt
    \\2. & A\hx 
    & \ge & a,
    \\3. & \hx  
    & \le  &  d, \mbox{ and }
    \\4. & B\hx
    & \le &(1+\eps)b+\beta.
  \end{array}
  \)
\end{thm}
\begin{pf}
  Note that the cost of $\x$ is a lower bound on $\opt$. Observe also
  that Step 1 does not change the space of integer solutions. 
  
  First we bound the cost of the solution $\hx'$ (to the restricted
  problem $\P'$).  Since $\x$ satisfies the KC inequalities for the
  specific set $F$, the definitions of $F$, $A'$, $b'$, and $d''$
  ensure that $\x'$ is a fractional solution of $\P'$.  By definition
  of $A^F$, the width of $\P'$ is at least 1.  Thus, the cost of
  $\hx'$ is $O(\ln (1+\alpha))$ times the cost of $\x'$, which is also
  $O(\ln (1+\alpha))$ times the cost of $\x$, and thus
  $O(\ln( 1+\alpha)\opt)$.
  
  Next we bound the cost of the final solution $\hx$.  The cost of
  $\hx$ is at most $1+\epsilon$ times the cost of $\x$, plus the cost
  of $\hx'$.  Thus, the cost of $\hx$ is $O(\ln( 1+\alpha)\opt)$.
  
  Next we verify that $\hx$ does not exceed the multiplicity
  constraints.  This is clear for the pinned variables: $\hx_j=d_j$
  for $j\in F$.  For the other variables ($j\not\in F$), we have
  \(\hx_j=\hx'_j \le \lceil  (1+\eps)d''_j \rceil =
\lceil (1+\eps)\x_j   \rceil  < \lceil (1+\eps)d'_j/(1+\eps) \rceil \le
d_j.\)

  Finally, $B\hx \le (1+\eps)b+\beta$
  follows from $B\x\le b$
  and 
  $\hx \le \lceil(1+\epsilon)\x\rceil$.
\qed \end{pf}

\begin{cor}
The integrality gap of the {\sc LP-KC} relaxation for CIP is $O(\ln (1+\alpha))$. 
\end{cor}

\subsection{Remarks on the necessity of the LP-KC relaxation}
Consider for simplicity that $d' = d.$ 
The algorithm starts with a ($1+\eps$)-relaxed solution $\x$ to LP-KC,
``pins'' $\hx_j=d_j$ for $j$ with $\x_j \ge d_j/(1+\eps)$, then uses
an existing bicriteria approximation algorithm to set the remaining
variables.  A natural question is whether the KC inequalities are
necessary.  Would it be enough to start with a fractional solution
$\x$ to the standard LP relaxation of the CIP?

If we do this, the analysis of the algorithm (as it stands) fails
because $\x'$ may no longer be a feasible solution to $\P'$.  (Indeed,
the problem $\P'$ may be infeasible with $d''$ defined as it is, or
even with $d''_j=d_j/(1+\eps)$.  To see this, consider the simple
example at the start of the section.)  This breaks the argument that
bounds the cost of $\hx$.

Perhaps the first fix that comes to mind is to modify the algorithm to
take $A'_{ij} = A_{ij}$ instead of $A'_{ij} = A^F_{ij}$ for
$j\not\in{}F$.  But this doesn't work because the resulting $\P'$ can
have width less than 1, worsening the approximation ratio.

Perhaps the second fix that comes to mind is to modify the algorithm
to, say, set $d''_j = d_j$ for $j\not\in F$, then solve $\P'$ from
scratch to obtain a (new) optimal fractional solution $\x''$.  In Step
7, the algorithm would pass that new fractional solution $\x''$ to
Theorem~\ref{theorem::bic} (instead of $\x'$) to compute $\hx'$.
Since the cost of $\x''$ is still a lower bound on $\opt$, it would
seem that we can again bound the cost of $\hx$ as desired.

The problem with this fix is that the new fractional solution $\x''$
can have $\x''_j > d_j / (1+\eps)$ for $j\not\in F$.  Indeed, it can
have $\x''_j = d_j$ for $j\not\in F$.  Thus, the rounded solution
$\hx'$ from Theorem~\ref{theorem::bic} could violate the multiplicity
constraints.

The natural work-around is to augment $F$ by adding any such $j$ to
$F$, then start over by returning to step 4 with the new $F$.  But, as
this process may repeat many times, it is not clear how one might
relate the cost of all the pinned variables to $\opt$.

\section{Open questions}  \label{sec::discussion}

Can one find in polynomial time an integer solution for CIP  with an additive
$1$ violation of the multiplicity constraints and logarithmic cost
guarantee with respect to the standard LP optimum (without KC
inequalities)?  We have shown this is possible for the case
$\max_j d_j=O(1)$.  Is there a faster (possibly greedy?)
$O(\ln{}m)$-approximation algorithm for CIP?

\begin{ack}
Thanks to Vijay Vazirani for a clarification on \cite{RajagopalanV93}.
Thanks to Lisa Fleischer and Aravind Srinivasan for comments on a
draft of this paper.
\end{ack}

\bibliography{/Users/stavros/papers/bibliography.bib}

\section*{Appendix}

\begin{pf} (of Lemma~\ref{lemma:folklore})
  We prove that $\hx$ is a $2L$-approximate solution with
  positive probability.  
  It suffices to prove that the probability
  that any of the following events happens is less than 1:
  \[
  (1)~c^{T} \hx > 2L c^{T} \x, ~\mbox{ or }~
  (2)~(\exists i)~~(A\hx)_i W/a_i < W.
  \]
  Note that $E[\hx] = x' = L\x$,
  so that by linearity of expectation
  \[ E[c^{T} \hx] = L E[c^{T} \x] = L\times(\optfrac_\infty), \] ~\mbox{ and }~
  \[ (\forall i)~E[(A\hx)_i W/a_i] = L (A\x)_i W/a_i \ge L W.
  \]
  By the Markov bound, the probability of (1) is at most $1/2$.
  
  Note that each $\hx_j$ can be thought of as a sum of independent
  random variables in $[0,1]$ (where we consider the fixed part,
  $\lfloor x'_j\rfloor$, to be the sum of $\lfloor x'_j\rfloor$
  variables each taking the value 1 with probability 1).
  Thus (by the choice of $W$) $(A\hx)_iW/a_i=\sum_j A_{ij}\hx_jW/a_i$
  is also a sum of independent random variables in $[0,1]$.  By a
  standard Chernoff bound \cite{RaghavanT87}, 
  \[\Pr[\,(A\hx)_iW/a_i \le (1-\eps) LW\,]
  ~<~ \exp(-\eps^2 LW/2).
  \]
  Taking $\eps$ such that $(1-\eps)L=1$,
  for the choice of $L$ in the rounding scheme,
  $\exp(-\eps^2 LW/2)\le 1/2m$.
  Thus, the above bound implies
  \[\Pr[\,(A\hx)_iW/a_i \le W\,] ~<~ 1/2m.\]
  Thus, by the naive union bound, the probability that
  (1) or (2) occurs is less than $1/2+m/2m = 1$.

  \medskip
  
  We have proven that the randomized rounding procedure returns a
  $2L$-approximate solution with positive probability.
\qed \end{pf}

\end{document}